\documentclass[prl,aps,twocolumn,floatfix,superscriptaddress,showpacs]{revtex4}

\usepackage{graphicx}
\usepackage{bm}

\begin{document}

\title{Selective coherence transfers in homonuclear dipolar coupled spin systems}

\author{Chandrasekhar Ramanathan}
\email{sekhar@mit.edu}
\affiliation{Department of Nuclear Engineering, Massachusetts Institute of Technology, Cambridge, Massachusetts 02139}
\author{Suddhasattwa Sinha}
\affiliation{Department of Nuclear Engineering, Massachusetts Institute of Technology, Cambridge, Massachusetts 02139}
\author{Jonathan Baugh}
\affiliation{Institute for Quantum Computing, University of Waterloo, Ontario, Canada N2L 3G1}
\author{Timothy F. Havel}
\affiliation{Department of Nuclear Engineering, Massachusetts Institute of Technology, Cambridge, Massachusetts 02139}
\author{David G. Cory}%
\affiliation{Department of Nuclear Engineering, Massachusetts Institute of Technology, Cambridge, Massachusetts 02139}%

\date{\today}

\begin{abstract}
Controlling the dynamics of a dipolar coupled spin system is critical to the development of solid-state spin based quantum information processors.  Such control remains challenging, as every spin is coupled to a large number of surrounding spins.  Here we demonstrate that in an ensemble of spin pairs it is possible to decouple the weaker interactions (weak coupling $\omega_D^w$) between different pairs and extend the coherence lifetimes within the two-spin system from 19 $\mu$s to 11.1 ms, a factor of 572.  This is achieved without decoupling the stronger interaction (strong coupling $\omega_D^S$) between the two spins within a pair.    An amplitude modulated RF field is applied on resonance with the Larmor frequency of the spins, with amplitude $\omega_1$, and frequency of the modulation matched to the strong coupling.  The spin pairs appear isolated from each other in the regime where the RF power satisfies $\omega_D^w \ll \omega_1 \ll \omega_D^S$.   
\end{abstract}

\pacs{03.67.Lx, 76.60.-k, 82.56.-b}

\maketitle

Nuclear spins feature prominently in most proposals for solid state quantum information processors.  They have the advantage of a simple and well defined energy level structure and they are normally well isolated from other degrees of freedom.  The challenge of using nuclear spins in solids is to obtain control over the multi-spin dynamics.  In a dielectric solid, the dominant interaction between the spins is the magnetic dipolar coupling.  Since the strength of the coupling between two spins is inversely proportional to the cube of the distance between them, a single spin is coupled to a large number of surrounding spins, and not just its immediate neighbors.  Therefore every desired gate is embedded in a complex, multi-body space and its dynamics has so far proven to be intractable.

Controlling the evolution of a dipolar coupled spin system has long been an important goal in solid state NMR, particularly for spectroscopic studies.  For example, the dipolar coupling has been effectively turned off using techniques such as spinning the sample rapidly at the magic angle ($\theta_m = \cos^{-1} (1/\sqrt{3})$) and a variety of multiple pulse techniques, which average the spatial and spin tensors of the coupling respectively, as well as a combination of these  \cite{Mehring,Haeberlen}.  

A very useful element of control would be to map the physical dipolar Hamiltonian of the spin system onto an effective interaction that has the form of only nearest neighbor couplings.  This would significantly simplify the implementation of accurate two-qubit operations in a many-qubit solid state spin-based quantum processor \cite{Cory,Wrachtrup,Yoshi,Suter}.  This restricted evolution is also necessary to avoid cross-talk between adjacent solid state quantum information processors in ensemble quantum computation \cite{Havel,Chuang}.   Without such control the gate fidelities achievable within a given processor element will be degraded due to leakage to other members of the ensemble.  Nearest neighbor mapping would also allow quantum simulations of many-body systems such as the Ising, XY or Heisenberg Hamiltonians in 1, 2 or 3 dimensions. 
The mapping envisioned here would have significant applications beyond quantum information processing.   For example, a nearest neighbor interaction could allow a more accurate determination of distances in NMR  structural studies, and could be used to perform sequential polarization transfers, such as along the backbone of a protein.

Here we report the first step towards the experimental realization of such a scheme for the special case of an ensemble of spin pairs, where the dipolar coupling between the spins within a pair is significantly larger than the coupling between spins on neighboring pairs.   We are able to extend the phase memory of the spin pairs by decoupling the pairs from each other, without decoupling the interaction between spins within a pair.
The control sequence consists of a simple amplitude modulated RF field, with the modulation frequency set to the desired dipolar coupling strength.

Our model system is an assembly of identical spin pairs with (strong) dipolar coupling
 $\omega_{D}^{S}$ and weaker couplings between spins on different pairs.
In a strong external magnetic field aligned along $\hat{z}$ the truncated secular dipolar Hamiltonian for this system is given by
\begin{equation}
 H_{d}  =  \frac{\hbar}{4} \sum_{i}\omega_{D}^{S}h_{12}^{ii} + \frac{\hbar}{4} \sum_{\alpha,\beta =1}^{2} \sum_{i\ne j} \omega_D^{ij\alpha\beta}h_{\alpha\beta}^{ij}
\label{eq:Hdip}
\end{equation}
where $h_{\alpha\beta}^{ij} = 2\sigma_{zi}^\alpha\sigma_{zj}^{\beta} - \sigma_{xi}^\alpha\sigma_{xj}^{\beta}  - \sigma_{yi}^\alpha\sigma_{yj}^{\beta}$, and $\omega_D^{ij\alpha\beta}$ is the coupling between spin $\alpha$ on pair $i$ and spin $\beta$ on pair $j$.  

The goal is to introduce a modulated RF field such that the effective Hamiltonian is restricted to just the isolated spin pairs (the first term in Eq.\ \ref{eq:Hdip}).  Our solution may be understood by viewing the RF field in the interaction frame of this coupling.  In the fully symmetric case the state $\sigma_x^1 + \sigma_x^2$ evolves as $\left(\sigma_x^1 + \sigma_x^2\right)\cos\left(\frac{3\omega_D^S}{2}t\right) + \left(\sigma_y^1\sigma_z^2 + \sigma_z^1\sigma_y^2\right) \sin\left(\frac{3\omega_D^S}{2}t\right)$, so we chose a RF  that has an amplitude modulation frequency $\omega_m = 3\omega_D^S/2$ and is given (in the lab frame) by
\begin{eqnarray}
 H_{\mathrm{mod}}(t) & = & \frac{\hbar\omega_1}{2} \cos\left(\frac{3\omega_D^S}{2} t\right) \times \nonumber \\ & & \hspace{0.2in} e^{i\frac{\omega_0 t}{2} \sum_i \sigma_z^i  } \left(\sum_i \sigma_x^i\right) e^{-i \frac{\omega_0 t}{2} \sum_i \sigma_z^i  }
\label{eq:Hrf}
\end{eqnarray}
where $\omega_0$ is the Larmor frequency of the spins.  The cosine amplitude modulation produces frequency sidebands at $\omega_0\pm\ 3\omega_D^S / 2 $.   Amplitude modulated pulses have previously been used in NMR for simultaneously irradiating multiple transitions in quadrupolar spin systems \cite{Vega-1} to create multiple quantum coherences in the regime where the RF power is significantly smaller than the strength of the quadrupolar coupling.

We illustrate a simple physical picture of the averaging process using a 3-spin case.  Consider the Hamiltonian (in the rotating frame) of a three-spin system, in which spins 1 and 2 are strongly coupled and spin 3 is weakly coupled to spins 1 and 2 ($\omega_D^S \gg \omega_D^w$), under the RF modulation
\begin{eqnarray}
H & = & \frac{\hbar\omega_D^{S}}{4}h_{12} +  \frac{\hbar\omega_D^w}{4}\left(h_{13}+h_{23}\right) + \nonumber \\ & &  \hspace{0.3in}  \frac{\hbar\omega_1}{2} \cos\left(\frac{3\omega_D^S}{2} t\right) \left( \sigma_x^1 + \sigma_x^2 + \sigma_x^3\right) \: \: .
\end{eqnarray}
 The time-dependent Hamiltonian in the interaction frame of the (1,2) pair interaction is given by 
\begin{equation}
\tilde{H}(t) = e^{-i(\omega_D^St) h_{12} / 4}\left(H - \frac{\hbar\omega_D^S}{4}h_{12}\right) e^{+i(\omega_D^S t) h_{12} / 4}
\end{equation}
The zeroth order average Hamiltonian \cite{Haeberlen} of this interaction frame Hamiltonian over a period $\tau = 4\pi / 3 \omega_D^{S}$ is 
\begin{equation}
\bar{H}^{(0)} = \frac{\hbar\omega_D^w}{2} \left(\sigma_z^1\sigma_z^3 + \sigma_z^2\sigma_z^3\right)+  \frac{\hbar\omega_1}{4} \left(\sigma_x^1 + \sigma_x^2\right) \: \: .
\end{equation}
The system can then be transformed into a second interaction frame via 
\begin{equation}
U' = \exp\left(\frac{i\omega_1 t}{4}\left(\sigma_x^1 + \sigma_x^2\right)\right) \: \: .
\label{eq:3Hinter2}
\end{equation}
Now, in this second averaging frame of $\bar{H}^{(0)}$ (Eq.\ \ref{eq:3Hinter2}), the first term, i.e.\ the residual dipolar couplings to spin 3, averages to zero over a cycle $\tau ' = 4\pi / \omega_1$.  Hence the second averaging of the couplings to spin 3 is efficient when $\omega_1 \gg 4\omega_D^{w} $, and the effective total system dynamics are generated by the (1,2) dipolar coupling (the Hamiltonian of the first frame transformation).  This picture provides the motivation for our approach, but the overall dynamics are more complicated than that suggested by the zeroth-order average shown above.  If $\omega_D^S$ is not significantly stronger than $\omega_1$, the higher order terms of the Magnus expansion become more important.  In addition, if the strength of the 1-3 and 2-3 couplings are different, additional two and three body terms appear in the Hamiltonian.

Further insight may be obtained by considering the energy level structure of an isolated dipole-coupled pair of spin-1/2 nuclei.  This has four energy levels, with three triplet
levels corresponding to a composite I = 1 system and a non-magnetic singlet with I = 0 \cite{Keller}.  A weakly coupled set of spin pairs will largely preserve this structure, but transitions between the singlet and triplet will no longer be forbidden due to the coupling between spins on different pairs.  Let $\omega_D^w$ represent the average strength of the weaker couplings.  Figure 1 illustrates how the spin pairs are decoupled from each other under the amplitude modulated (AM) RF irradiation.    If $\omega_{m} = 3\omega_{D}^{S}/2$, the AM irradiation simultaneously
drives transitions  $|11\rangle- |00\rangle \leftrightarrow
|10\rangle+|01\rangle$  and $|11\rangle-|00\rangle \leftrightarrow
|00\rangle+|11\rangle$ of the triplet manifold. The rate at which
these transitions are driven depend on the strength of the
modulation field, $\omega_{1}$.  If $\omega_{1} \gg
\omega_{D}^{w}$, the two triplet manifolds are decoupled from
each other and the pairs are isolated from each other.  
However, if the RF power is increased further such that $\omega_{1} \ge \omega_{D}^{S}$, the triplet sub-space structure gets destroyed as
the strong coupling between the spins within the pair is decoupled. 

Thus,
our scheme works in the regime where $\omega_m = 3\omega_D^S/2$ and $\omega_{D}^{S} \gg \omega_{1}\gg \omega_{D}^{w}$.  Not surprisingly these are exactly the same conditions as obtained in the previous section. 
Intuitively, the RF modulation allows us to move into an interaction frame that is moving with the magnetization of the dipole coupled spin pair.   The experiment bears some similarities to the spin-1 decoupling experiments originally proposed by Pines and coworkers \cite{Pines-1,Pines-2}.  In fact they suggest that their method could be used to decouple a heteronuclear spin from a pair of identical spins.  However, the scheme presented here goes further, and permits a coherent evolution of the isolated spin pair while decoupling the pairs from each other.
\begin{figure}
\centering\centering{\resizebox{90mm}{!}{
\includegraphics{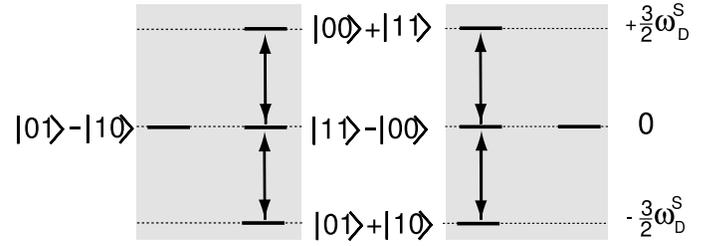}}}
\caption{\label{fig:fig1b} Two pairs of strongly coupled
spin--$\frac{1}{2}$ systems with each pair decomposed into its
singlet and triplet manifolds (in the rotating frame). The triplet manifolds are weakly
coupled to each other while the singlet manifolds do not interact.  
 }
\end{figure}
 \begin{figure}
\centering\centering{\resizebox{60mm}{!}{
\includegraphics{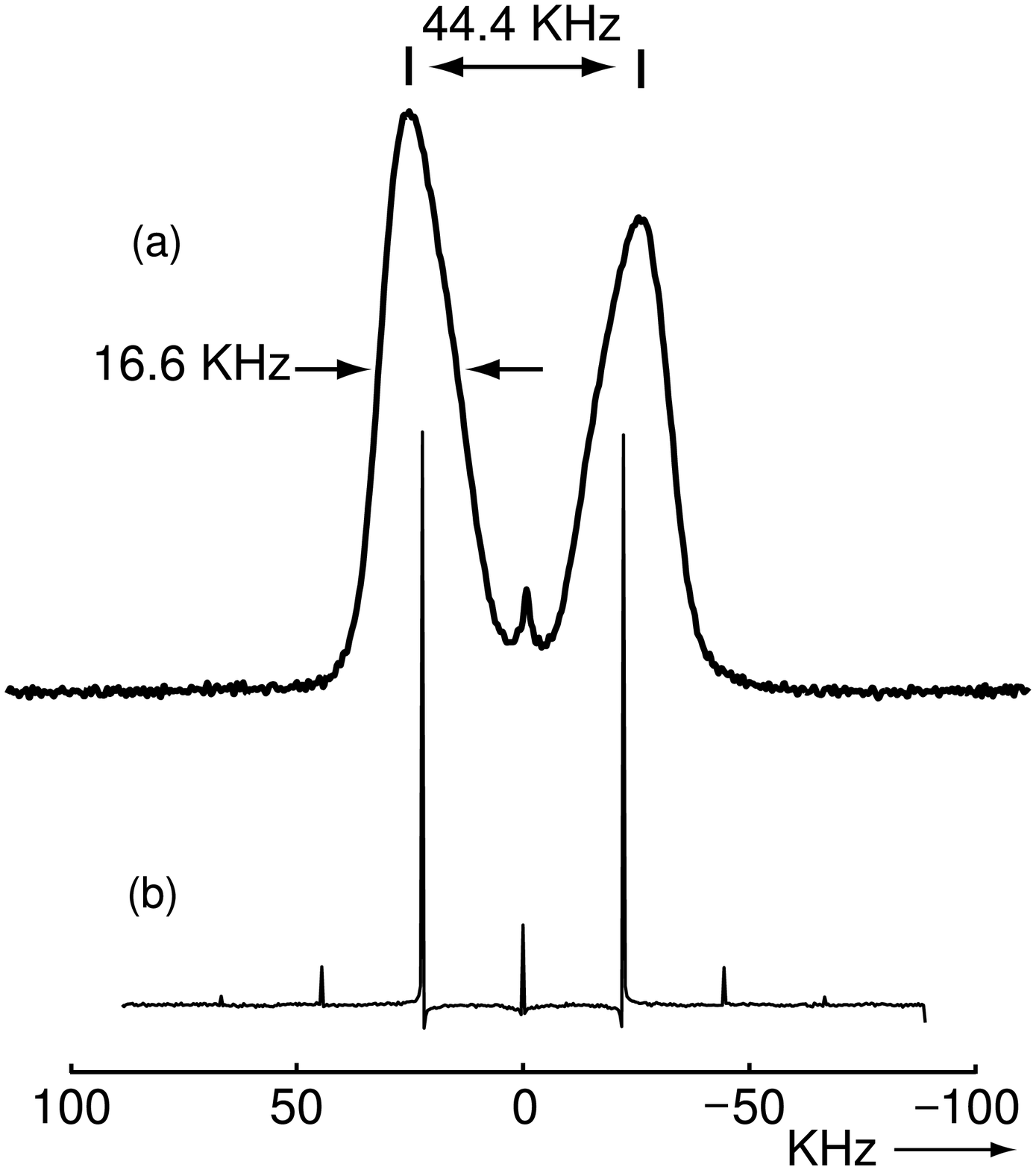}}}
 \vspace*{0.2in}
\centering\centering{\resizebox{60mm}{!}{
\includegraphics{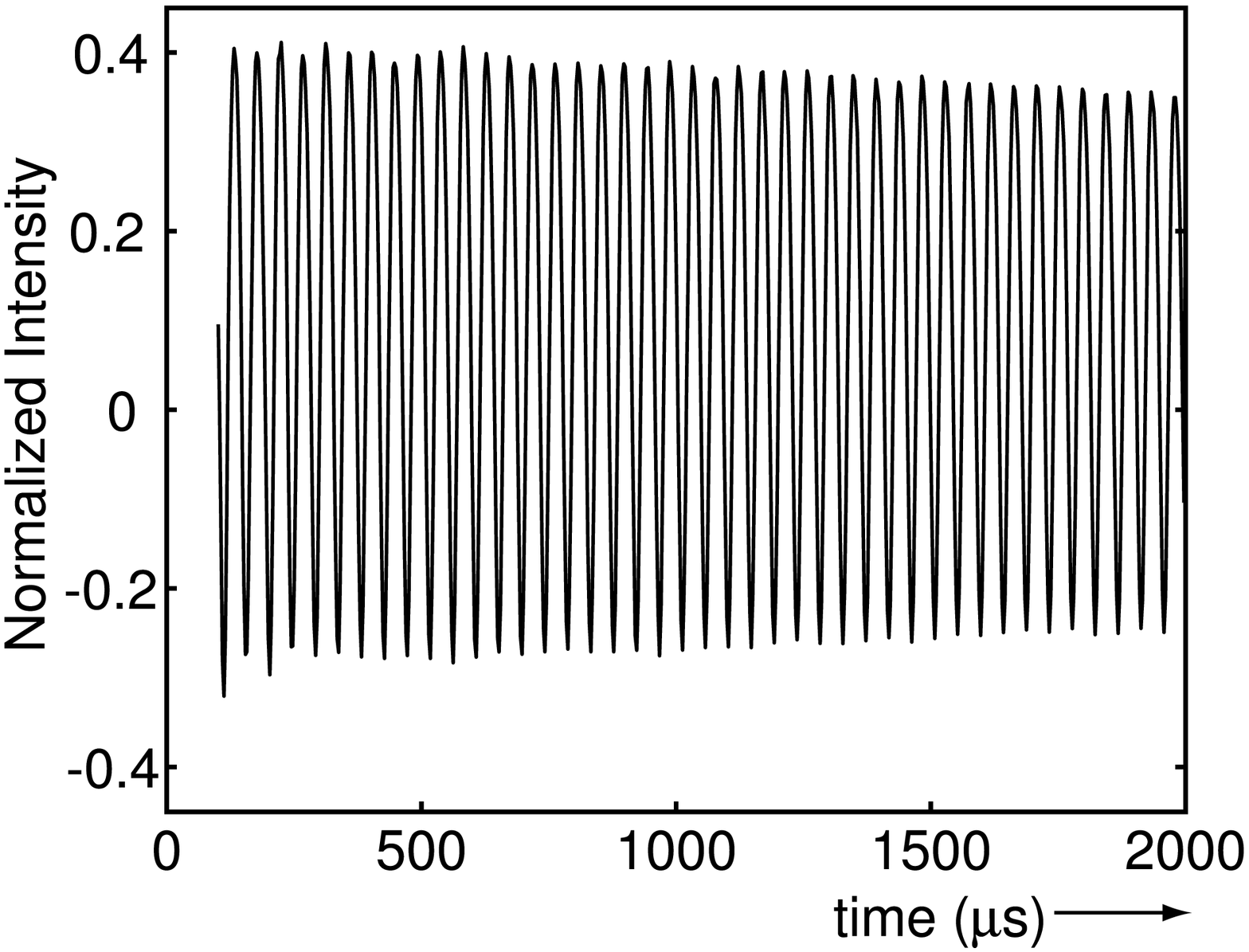}}}
\caption{(a) The proton spectrum of a single
crystal of gypsum in the [010] orientation  (Pake doublet). The splitting between
the peaks corresponds to $3\omega_{D}^{S}/2\pi = 44.4 $
KHz. Each peak is broadened due to weak dipolar coupling to the
other water molecules. Therefore, $3\omega_{D}^{w}/2\pi \sim 16.4 $
KHz (b) The narrowing down of peaks under the modulation sequence
with RF amplitude $\omega_{1} =3\omega_D^S/4 $ .  (c) Evolution of the single spin $\sigma_x$ terms under the
modulation sequence.}
\label{fig:spectra}
\end{figure}

It is also useful to move into the interaction frame of the RF modulation.
The zeroth-order average Hamiltonian \cite{Waugh} of an isolated spin pair in the interaction frame of the RF modulation is
\begin{eqnarray}
\bar{H}_{d}^{(0)} & = &  -\frac{\hbar}{8} \omega_{D}^{S}\left(2\sigma_{x}^1 \sigma_{x}^2
 - \sigma_{y}^1 \sigma_{y}^2 - \sigma_{z}^1 \sigma_{z}^2\right) + \nonumber \\  & & \frac{3\hbar}{8}\sum_{i}\omega_D^{S} J_0 \left( \frac{4\omega_{1}}{\omega_D^S}\right)
\left(\sigma_{z}^1 \sigma_{z}^2 - \sigma_{y}^1 \sigma_{y}^2\right)  \: \: .%
\label{eq:interactionH}
\end{eqnarray}
where the average has again been performed over one period of the amplitude modulation ($t = 2\pi / \omega_m = 4\pi / 3\omega_D^S$).
Starting from the equilibrium state where the spins are along the external magnetic field, a collective $\pi /2$ rotation of the spins places a spin pair in the initial state $\sigma_x^1 +
\sigma_x^2$.  This state commutes with the first term of the interaction Hamiltonian shown above,
and the effective evolution is only due to the second term $\sigma_z^1 \sigma_z^2 - \sigma_y^1 \sigma_y^2$.  The set of operators, $(\sigma_x^1 +
\sigma_x^2,~\sigma_z^1 \sigma_y^2 + \sigma_y^1
\sigma_z^2,~\sigma_z^1 \sigma_z^2 - \sigma_y^1 \sigma_y^2)$ form
a subalgebra under the commutator that is isomorphic to the Cartesian subalgebra
$(\sigma_x,\sigma_y,\sigma_z)$.  Thus the strongly coupled spins oscillate between the single spin
state $\sigma_x^1 + \sigma_x^2$ and the two spin state $\sigma_z^1 \sigma_y^2 +
\sigma_y^1 \sigma_z^2$.   If the first term of the Hamiltonian in Eq.\ \ref{eq:interactionH} were absent, this scheme would map onto a nearest neighbor interaction, and as long as the initial state of the spin pairs was within this subspace, leakage out of the subspace would be substantially suppressed.  However, in the current scheme, the initial state should be both within the subspace and commute with the first term of Eq.\ \ref{eq:interactionH}.  For an ensemble of spin pairs, only the collective $\sigma_x$ state satisfies these conditions.

Gypsum (CaSO$_4 \cdot2$ H$_2$O) was taken as a prototypical system for a
weakly interacting ensemble of identical spin pairs.  The protons in the waters of crystallization
comprise the strongly coupled spins.  The coupling between protons on different water
molecules is significantly smaller than that between protons in the same molecule.
A unit cell of gypsum has four water molecules, with two pairs in two inequivalent sites.
When the external magnetic field is applied along the [010] orientation,
the dipolar splitting at the inequivalent water sites coincide and
a Pake doublet is observed (see Fig.\ \ref{fig:spectra})  in a one-pulse experiment \cite{Pake}.
In this orientation the strong dipolar coupling between protons in the water molecule is $\omega_D^S / 2\pi$ = 14.8 kHz, and the mean coupling between protons on different water molecules is $\omega_D^w / 2\pi$ = 5.5 kHz.

\begin{figure}
\centering\centering{\resizebox{80mm}{!}{
\includegraphics{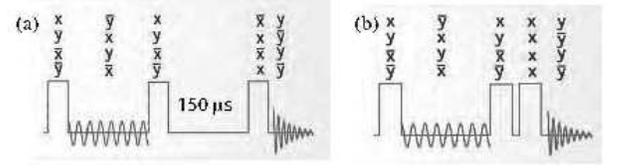}}}
\caption{(a) pulse sequence used to read out the $\sigma_x^1 + \sigma_x^2$ terms.  Following the modulation pulse, a $\pi / 2$ pulse is applied to rotate the $\sigma_x$ terms to $\sigma_z$.  During the 150 $\mu$s interval (much shorter than T$_1$) all terms other than the $\sigma_z$ decay.  A $\pi / 2$ pulse is then used to monitor $\sigma_z$.    (b) pulse sequence used to read out the $\sigma_y^1\sigma_z^2 + \sigma_z^1\sigma_y^2$ term.  Following the modulation two back to back $\pi / 2$ pulses act as a double quantum filter to suppress the single spin $\sigma_x$ terms.  A four step phase cycle is necessary to implement the filter.}
\label{fig:pulses}
\end{figure}
\begin{figure}
\centering\centering{\resizebox{70mm}{!}{
\includegraphics{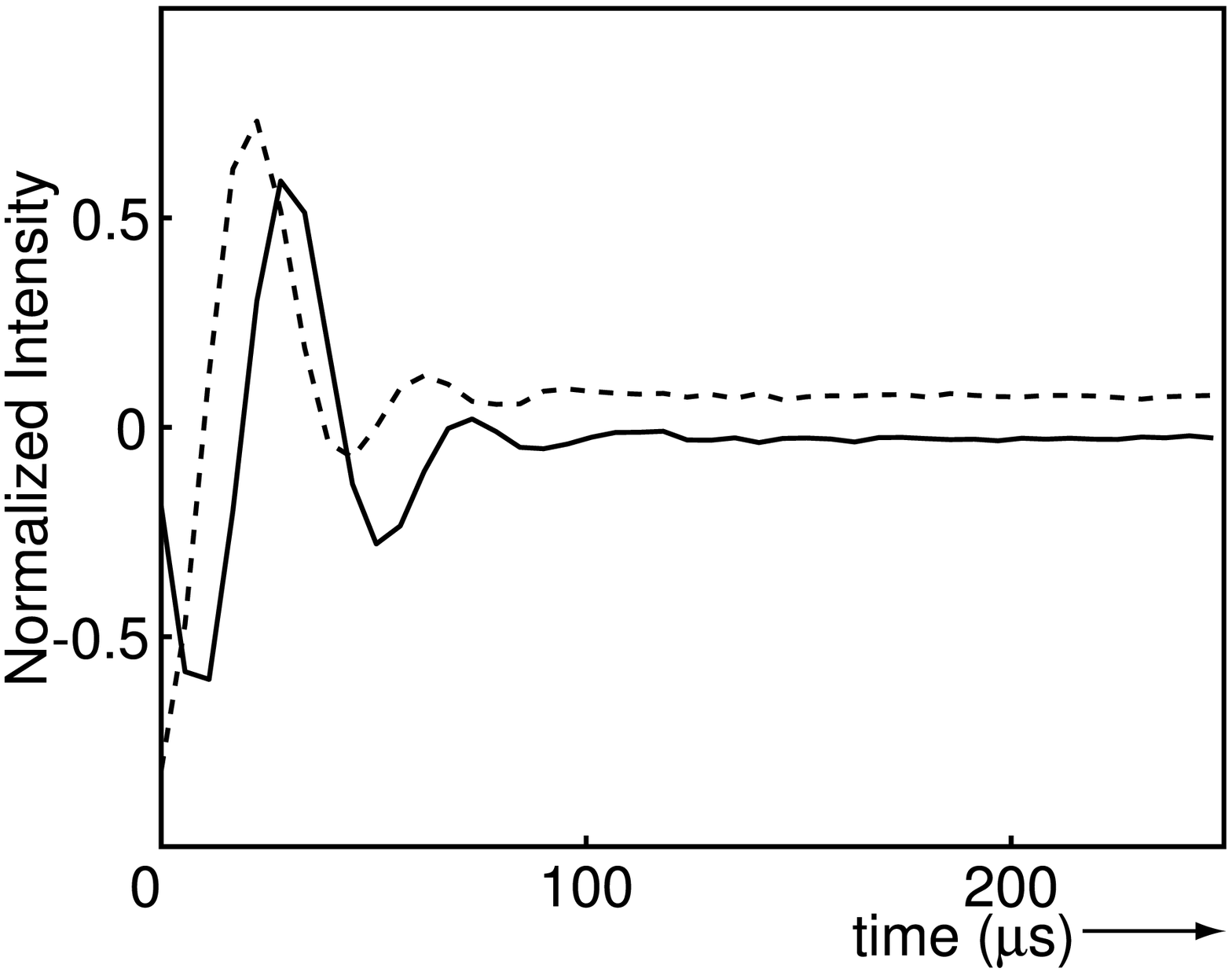}}}
\centering\centering{\resizebox{70mm}{!}{
\includegraphics{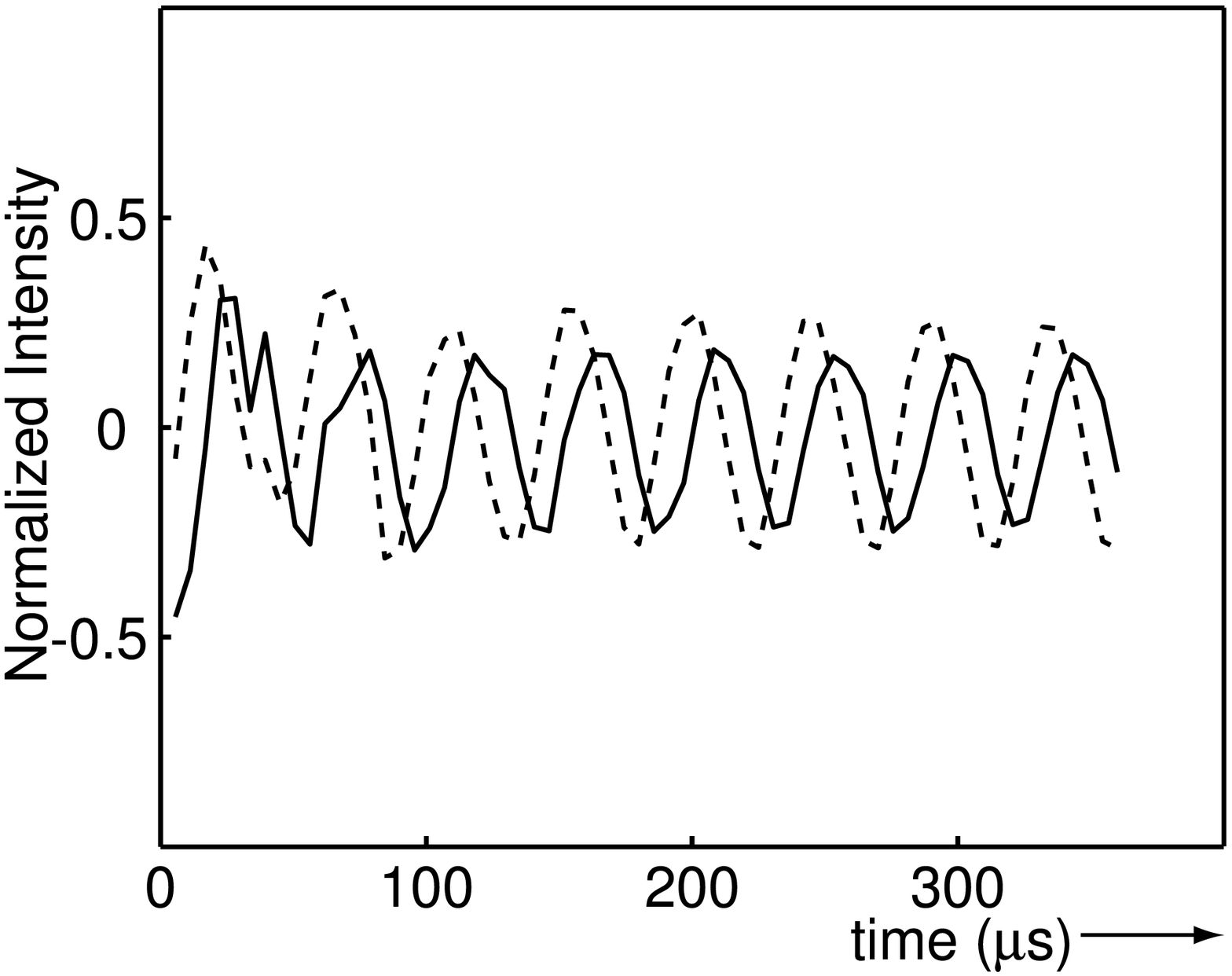}}}
\caption{The dashed line shows the $\sigma_x^1 + \sigma_x^2$ and
the solid line shows the $\sigma_z^1 \sigma_y^2 + \sigma_y^1 \sigma_z^2$ terms. (i) Under the dipolar Hamiltonian evolution,
the above terms evolve into unobservable higher order spin correlations within
100 $\mu$s. (ii) Under the modulation sequence, the terms oscillate 90 $\deg$ 
out-of-phase with each other for up to 360 $\mu$s without any
significant attenuation in amplitude. In this case the RF amplitude $\omega_{1} =3\omega_D^{S}/4$.}
\label{fig:coherencetransfer}
\end{figure}

The experiments were
carried out at room temperature at 7.1 T ($^1$H 300 MHz) using a
Bruker Avance spectrometer on a 1 mm$^3$  single crystal 
of gypsum in the [010] orientation. The length of the $\pi/2$ pulse
used was 1.67 $\mu$s.  The experiment was repeated as the duration of the AM RF was varied from 100 $\mu$s to 2.9 ms with an increment of 5.5 $\mu$s.   The signal intensity was Fourier transformed with respect to the length of the modulation pulse to yield the spectrum shown in Fig.\ \ref{fig:spectra}(b).  Fig.\ \ref{fig:spectra}(c) shows the observed $\sum_i \sigma_x^i$ terms plotted against the length of the modulation pulse \cite{notes}.   A dramatic narrowing of the spectral line is observed in the experiment.  The effective T$_2$ of the spins under the modulation is $ 11.1$ ms which corresponds to a linewidth of 29 Hz.  This is a factor of 572 times smaller than the 16.6 kHz width of a single line of the Pake doublet.  

In order to demonstrate that the spin pair continues to undergo a coherent evolution, we performed a second series of experiments to specifically filter out and separate the $\sigma_x^1 + \sigma_x^2$ and the  $\sigma_y^1 \sigma_z^2 + \sigma_z^1 \sigma_y^2$ terms.   The two experiments are shown in Figure \ref{fig:pulses}.  Figure \ref{fig:coherencetransfer}(a) shows the coherence transfer using the dipolar Hamiltonian while Fig.\ \ref{fig:coherencetransfer}(b) shows the coherence transfer under the action of the modulation sequence.  Under the dipolar coupling the interactions with distant spins rapidly generate higher order spin correlations, and there is a strong damping of the oscillation between the single spin and the two-spin terms.  However, under the modulation sequence this oscillation is seen to extend out significantly farther.   Thus the observed line-narrowing is not a form of spin-locking of the single spin terms, as occurs under strong RF irradiation, but is due to the selective decoupling of the weaker interactions between spins on different pairs.

In conclusion,  we have demonstrated that it is possible to restrict the evolution of a dipolar coupled spin network to a much smaller subspace of the system Hilbert space.  This restriction allows us to significantly extend the phase coherence times for selected states.  The scheme developed works for a system consisting of an ensemble of spin pairs, where the coupling between spins in the same pair is stronger than the coupling between spins on different pairs.  

\begin{acknowledgments}

We would like to thank  Nicholas Boulant, HyungJoon Cho, Thaddeus Ladd (Stanford University) and Professor Raymond Laflamme (Institute for Quantum Computing, Waterloo, Canada), for helpful discussions and the ARO, DARPA , ARDA, NSERC and CIAR for financial support. 

\end{acknowledgments}

\bibliography{revdip}

\begin{thebibliography}{15}
\bibitem[{not()}]{notes}
  \bibinfo{note}{The signal intensity was measured using a solid echo
  sequence. The amplitude of the 15th point on the acquired data was then
  either plotted against the duration of the RF modulation or Fourier
  transformed to yield a spectrum}.
\end{thebibliography}

\end{document}